\RequirePackage{fix-cm}
\documentclass[smallextended]{svjour3}       
\smartqed  
\usepackage{graphicx}
\usepackage{amsfonts,amssymb,amsmath}
\usepackage{epsfig}
\usepackage{mathrsfs}
\usepackage{amssymb}
\usepackage{graphicx}
\usepackage{amsmath}
\begin{document}

\title{Analysis of $R^p$ inflationary model as $p\geqslant 2$
}


\author{Lei-Hua Liu}


\institute{Lei-Hua Liu\at
              Institute for Theoretical Physics, Spinoza Institute and
the Center for Extreme Matter and Emergent Phenomena (EMME$\Phi$), Utrecht University,
Buys Ballot Building
Princetonplein 5, 3584 CC Utrecht, the Netherlands \\
              Tel.: +31 (30) 253 5907\\
              \email{L.Liu1@uu.nl}           
           \and
}

\date{Received: date / Accepted: date}

\maketitle

\begin{abstract}
We study the $R^p$ inflationary model of \cite{Muller:1989rp} for $p>2$ using the result of Ref.~\cite{Motohashi:2014tra}. After reproducing the observable quantities: the power spectral index $n_s$, its corresponding running $\alpha=\frac{dn_s}{d\ln(k)}$ and the tensor to scalar ration $r$ in terms of e-folding number $N$ and $p$, we show that $R^p$ inflation model is still alive as $p$ is from $2$ to $2.02$. In this range, our calculation confirms that $n_s$ and $r$ agree with observations and $\alpha$ is of order $10^{-4}$ which needs more precise observational constraints. We find that, as the value of $p$ increases, all $n_s$, $r$ and $|\alpha|$ decrease. The precise interdependence between these observables is such that this class of models can in principle be tested by the next generation of dedicated satellite CMB probes.

\keywords{Inflation, consistency relations, constraints}
 \PACS{98.80.-k, 98.80.Bp, 98.80.Es.}
\end{abstract}

\section{Introduction}
\label{intro}
Currently, $\Lambda CDM$ model is the paradigm providing a consistent explanation for the acceleration expansion of universe, the formation of large scale structure, cold dark matter and even for the most mysterious dark energy. However, it still suffers from the horizon, flatness, homogeneity and so-called magnetic monopole problems. When supplementing with a scenario of inflation~\cite{Starobinsky:1980te,Guth:1980zm,Linde:1981mu}, these problems can be solved elegantly. In this framework, we need an additional scalar field called inflaton to trigger the inflationary period. In order to relate to the standard model (SM), there are lots of SM particles can be produced from preheating process~\cite{Kofman:1997yn}. Following Occam's razor principle, if we embed the inflaton field in the a Higgs sector, one can rule out this model due to its large value of $r$. However, it is still alive by introducing a non-minimal coupling between Ricci scalar and Higgs field $h$~\cite{Bezrukov:2007ep}.

 By introducing the non-minimal coupling, a notable alternative called modified gravity is proposed whose effective scalar fields are called {\it scalarons} -- which can be generated by {\it e.g.} quantum effects. To be more precise, generic quantum fluctuations from matter around the Plank energy scale can generate higher derivative local gravitational operators in the effective action. One class of these operators are a function of the Ricci scalar $R$ denoted by $f(R)$. By introducing a Lagrange multiplier and transforming into the Einstein frame, these operators can play a role of {\it inflaton}. The most simple and successful model was proposed by Starobinsky~\cite{Starobinsky:1980te} whose Lagrangian is $\mathfrak{L} =\frac{M_{\rm P}^2}{2} R - a R^2$, where $M_P=\frac{1}{\sqrt{8\pi G}}$ is the reduced Plank mass and $a\sim 10^9$. Its predictions still agree very well with the current observational constraints. \cite{Ade:2015xua}.

The high accuracy of $R^2$ inflation model motivates us to propose various modified gravity models to mimic the evolution of universe model under the framework of $f(R)$ gravity, {\it i.e.} by choosing suitable form for $f(R)$ including non-linear terms, the accelerating expansion without cosmological constant can be reproduced \cite{Carroll:2003wy,Capozziello:2002rd,Sotiriou:2006qn}. These $f(R)$ gravity models even can pass Solar sysmem tests \cite{Hu:2007nk,Nojiri:2007cq}. Further, these models unify the inflation and cosmic acceleration \cite{Nojiri:2003ft,Nojiri:2010wj}. Models such as $f(R)=R+F(R)$, where $f(R)$ and $F(R)$ are functions of Ricci scalar, are still suffering from the singularity problem since the scalar mass and Ricci scalar are divergent in the very beginning of the Universe \cite{Starobinsky:2007hu,Tsujikawa:2007xu}.  This crucial problem can be solved by inserting the term $R^2$ into $f(R)$ \cite{Appleby:2009uf}, similar idea was proposed for solving the finite time singularity \cite{Bamba:2008ut,Nojiri:2008fk}. The dynamical part for inflation is the same, the deviation will appear in the reheat phase dominated by the kinetic term of scalaron \cite{Motohashi:2012tt}. Meanwhile, it enhances the tensor power spectrum \cite{Nishizawa:2014zra}. Inspired by the ultraviolet complete theory of quantum gravity, Ref.~\cite{Huang:2013hsb} considers that a polynomial f(R) inflation model where $f(R)=R+\frac{R^2}{6M_P^2}+\frac{\lambda_n}{2n}\frac{R^n}{(3M^2)^{n-1}}$, they find that $R^n$ is exponentially suppressed. Also nearly the Starobinsky inflationary model, Ref. \cite{Sebastiani:2013eqa} generalizes to the class of inflationary scalar potentials $V(\sigma)\propto \exp(n\sigma)$, in which $n$ can be included in $F(R)$ where it is the polynomial of Ricci scalar to present various models of $F(R)$ gravity. In the light of conformal transformation from Jordan frame to Einstein frame, one can reconstruct viable inflationary model  \cite{Odintsov:2018ggm,Kuiroukidis:2017dyz,Oikonomou:2018npe}. Inspired by $\alpha$ attractor \cite{Kallosh:2013hoa,Kallosh:2015lwa,Galante:2014ifa}, even one can reconstruct the $f(R)$ gravity from $\alpha$ attractor \cite{Miranda:2017juz}.

Thus, in order to find a most economical generating theory of $R^2$ inflation without suffering from the singularity problem, the so-called $R^p$ inflation was proposed \cite{Muller:1989rp,Gottlober:1992rg}. $R^p$ inflation model could also give the correction to $R^2$ inflation \cite{Codello:2014sua,Ben-Dayan:2014isa,Rinaldi:2014gua}. Together with this framework, inflation can also be reproduced in higher dimensions through compactification \cite{Nakada:2017uka,Ketov:2017aau}. From perspective of dark energy, the constraint for $R^p$ inflation model can also be given \cite{Geng:2015vsa}. All of these models give the value of $r$ that is little larger when compared to $R^2$ inflation model as requiring $p<2$. Here, we consider the case of $p\geqslant 2$ in $f(R)=R+R^P$ and then we study the scalar spectral index $n_s$, the tensor-to-scalar ratio $r$ and the running of scalar spectral index $\alpha=\frac{dn_s}{d\ln k}$ in order to compare with current observational constraints.

The organization of the paper is as follows. In Section 2, we present the $R^p$ inflation model and the slow-roll approximation is achieved. In Section 3, results are presented according to the consistency relation. Section 4, we give our main results and conclusion.

\section{The model}

Recently, Ref.~\cite{Motohashi:2014tra} derives a consistency relation in $R^p$ inflationary model. Using their results, we study this model as $p\geq 2$. Firstly, we recap how to get this model from $f(R)$ gravity. The effective action comes from the most economical generalization of $R^2$ inflation,

\begin{equation}
S=\int d^4 x\sqrt{-g}\frac{M_p^2}{2}f(R)
\,,
\label{Jordan action}
\end{equation}
where $M_P=(8\pi G)^{-1/2}$ is the reduced Plank mass and $f(R)=R+\lambda R^p$ and $\lambda$ is a constant and $[\lambda]=2-2p$.

Next we need to proceed to investigate inflation governed by action~(\ref{Jordan action}).  This action is on-shell equivalent to
\begin{equation}
S=\int d^4 x\sqrt{-g}\frac{M_p^2}{2}\left[f(\Phi)+\omega^2(R-\Phi)\right]
\,,
\label{Jordan action 2}
\end{equation}
where $f(\Phi)=\Phi+\lambda \Phi^p$, $\Phi$ is a real scalar field (dubbed scalaron in \cite{Starobinsky:1980te}) and $\omega=\omega(x)$ is a Lagrange multiplier (constraint) field. Upon varying the acition~(\ref{Jordan action 2}) and solving the resulting equation we can obtain,
\begin{equation}
f'(\Phi)-\omega^2=0
\,,
\label{Jordan action 3}
\end{equation}
where

\begin{equation}
f'(\Phi)=\frac{df}{d\Phi}\equiv F(\Phi)
\,,
\label{Fphi}
\end{equation}

Inserting action (\ref{Jordan action 3}) into action (\ref{Jordan action 2}) which is on-shell equivalent to (\ref{Jordan action}), we obtain:
\begin{equation}
S=\int d^4 x\sqrt{-g}\frac{M_p^2}{2}\left[f(\Phi)+F(\Phi)(R-\Phi)\right]
\,.
\label{Jordan action 4}
\end{equation}
Note that $\Phi$ as a scalar field is non-minimally coupled to gravity via the $F(\Phi)R$. Next step is to transform the Jordan frame (action (\ref{Jordan action 4})) into Einstein frame by a conformal transformation $g_{\mu\nu}=\Omega^2(x)g_{\mu\nu}^E$, where $\Omega=\Omega(x)$ is a some specific local function. Upon this transformation is executed, action (\ref{Jordan action 4}) becomes
\begin{eqnarray}
S=\int d^4 x\sqrt{-g_E}\frac{M_p^2}{2}\bigg[\Omega^{2}F
   \bigg(\!R_E-6g^{\mu\nu}_E \frac{\nabla_\mu^E \nabla_\nu^E\Omega}{\Omega}
\bigg)-\Omega^4\Big(F(\Phi)\Phi-f\Big)
\bigg].
\label{Einstein action 1}
\end{eqnarray}
By choosing an appropriate function for conformal transformation,
\begin{equation}
\Omega^{2}=\frac{1}{F(\Phi)}
\,,
\label{Omega vs Phi}
\end{equation}
and then partially integrating the second term in the bracket of action (\ref{Einstein action 1}), and dropping the boundary term, the action becomes,
\begin{equation}
S=\int d^4 x\sqrt{-g_E}\left[\frac{M_{\rm P}^2}{2}R_E
 -3M_{\rm P}^2g^{\mu\nu}_E\frac{\nabla_\mu^E\Omega \nabla_\nu^E\Omega}{\Omega^2}
 -\frac12\Omega^4(F\Phi-f)\right]
\,.
\end{equation}
The higher gravitational operator has disappeared, but a new dynamical scalar field appeared named scalaron field. Note that scalaron is of non-canonical kinetic form, it can be changed into the canonical term by a simple transformation to {\it Einstein frame},
 \begin{equation}
\phi_E=-\frac{M_P}{2}\sqrt{6}\ln\big(\Omega(\Phi)\big)
\,,
\label{transform to Einstein frame}
\end{equation}
where the mapping between these two fields is chosen such that $\phi_E=0$ as $\Omega=1$. The field $\phi_E$ can have the opposite sign since the resulting potential would be of mirror symmetry around $\phi_E=0$ of the potential from (\ref{transform to Einstein frame}). With this in mind, action (\ref{transform to Einstein frame}) finally becomes,
\begin{equation}
S=\int d^4 x\sqrt{-g_E}\left[\frac{M_{\rm P}^2}{2}R_E-\frac{1}{2}g^{\mu\nu}_E\partial_\mu\phi_E\partial_\nu\phi_E
             -V_E(\phi_E)\right]
\,,
\label{Einstein action 2}
\end{equation}
where $V_E(\phi_E)$ denotes the Einstein frame potential,
\begin{equation}
V_E(\phi_E)=\frac{M^2_P}{2}\frac{F\Phi-f}{F^2}
\,.
\label{Einstein frame potential}
\end{equation}
In light of Eqs.(\ref{Fphi}), (\ref{Omega vs Phi}) and (\ref{transform to Einstein frame}), we can write down the formula for $F(\phi_E)$,
 \begin{equation}
F(\phi_E)=\exp(\sqrt{\frac{2}{3}}\frac{\phi_E}{M_P})=1+(p-1)\lambda \Phi^{p-1}
\,.
\label{Fphi1}
\end{equation}
This equation defines the mapping $\phi_E=\phi_E(\Phi)$. From a theoretical perspective of Ref.~\cite{Starobinsky:1980te}, gravity is an effective field theory and every effective field theory can be quantized, here these two dynamical quantized fields are inflaton $\phi_E$ and graviton $g_{\mu\nu}^E$. In the proceeding part, we will discuss the dynamics of classical fields (condensate state of quantum state from macroscopic perspective) and the (tree level) dynamics of first order quantum perturbations.

\subsection{Background dynamics and Cosmological perturbations}

Action (\ref{Einstein action 2}) is usually considered as driving inflation and it provides the quantum field $\hat\phi_E$ for supporting large expectation values for inflation. Assuming that the field is approximately homogeneous with respect to some space-like hypersurface, it can be decomposed into the inflaton part and a small perturbation as follows,
\begin{equation}
  \hat \phi_E(x) = \phi_{E0}(t) + \hat \varphi_E(x)
 \,,\quad
\phi_{E0}(t) =\langle \hat \phi_E(x) \rangle \equiv {\rm Tr}\left[\hat\rho(t)\hat \phi_E(x) \right]
\,.
\label{inflaton and perturbaiton}
\end{equation}
Similarly, the tensor metric can also be decomposed into two parts (in Einstein frame),

 \begin{eqnarray}
  \hat g_{\mu\nu}(x) &=& g^b_{\mu\nu}(t)+  \delta \hat g_{\mu\nu}(x)
 \,,\\
 g^b_{\mu\nu}(t) &=&\langle \hat g_{\mu\nu}(x) \rangle = {\rm diag}\left(-1,a_E^2(t),a_E^2(t) , a_E^2(t)\right)
\,,
\label{graviton and perturbaition}
\end{eqnarray}
 the form of $\delta g_{\mu\nu}=a^2 h_{\mu\nu}(x)$
is written in conformal time, so the scale factor should be a function of conformal time,
$a=a(\tau)$. Alternatively, and probably better, already at this state one can fix the gauge
to the traceless-transverse gauge. In this gauge tensor perturbations are given
by the spatial part of the metric perturbation, and thus can be written in the form,
$\delta g_{ij}(t,\vec x)=a^2(t) h_{ij}(t,\vec x)$, with $\delta_{ij} h_{ij}=0$ and
$\partial_i h_{ij}=0$.

The dynamical equation for the inflaton condensate in the background of an expanding universe is governed by,
\begin{equation}
 \ddot \phi_{E0}(t) + 3H_E\dot\phi_{E0}(t)+\frac{dV_E}{d\phi_{E0}}=0
\,,
\label{EOM phi0}
\end{equation}
where $H_E(t)$ is the Hubble parameter in Einstein frame and we have neglected the backreaction from quantum fluctuations. Since inflaton drives inflation, the universe's dynamics is governed by Friedmann (FLRW) equations,
 \begin{eqnarray}
 H_E^2 &\equiv& \left(\frac{\dot a_E}{a_E}\right)^2
    = \frac{1}{3M_{\rm P}^2}\left(\frac{\dot \phi_{E0}^2}{2}+V_E(\phi_{E0})\right)
\label{Friedmann 1}\\
 \dot H_E
    &=& -\frac{\dot \phi_{E0}^2}{2M_{\rm P}^2}
    \label{Friedmann 2}
\,,
\end{eqnarray}
where $a_E$ is scale factor in Einstein frame and $\dot H_E = dH_E/dt$. The equation of motion (EOM) for the inflaton perturbation and graviton perturbation are governed by,
\begin{eqnarray}
  \left(\frac{d^2}{dt^2} + 3H_E\frac{d}{dt} + \frac{k^2}{a_E^2}+\frac{d^2V_E}{d\phi_{E0}^2}\right)\varphi(t,k) = 0\\
  \left(\frac{d^2}{dt^2} + 3H_E\frac{d}{dt} + \frac{k^2}{a_E^2}\right)h(t,k) = 0
\,,
\label{EOM mode function}
\end{eqnarray}
 where $\phi(t,k)$ and $h(t,k)$
are the Fourier modes of $\hat {\phi}_E(t,\vec x)$ and $\hat h_{ij}(t,\vec x)$ and here
$k=\|\vec k\|$. After adopting the zero curvature gauge which means that the spatial scalar graviton perturbations vanish, then we obtain the scalar and tensor spectra:
\begin{eqnarray}
  \Delta^2_s(k) &=&\Delta^2_{s*}\left(\frac{k}{k_*}\right)^{n_s(k)-1}=\frac{k^3}{8\pi^2\epsilon_EM_{\rm P}^2}|\varphi(t,k)|^2
\nonumber\\
 \Delta^2_t(k) &=& \Delta^2_{t*}\left(\frac{k}{k_*}\right)^{n_t(k)}= \frac{2k^3}{\pi^2M_{\rm P}^2}|h(t,k)|^2 = 16\epsilon_E \Delta^2_s
\,,
\label{observer spectra}
\end{eqnarray}
where $k_*=0.05~({\rm Mpc})^{-1}$ or $k_*=0.002~({\rm Mpc})^{-1}$ and it is a fiducial comoving momentum, $\Delta^2_{s*}\equiv A_s$ and $\Delta^2_{t*}$ are the amplitude of scalar and tensor spectra evaluated at $k=k_*$
and $n_s$ and $n_t$ are the scalar and tensor spectral indices, respectively. These two spectra are obtained in slow roll approximation.
Then by performing the canonical quantization and choosing the Bunch-Davies vacuum, we obtain identical expressions for $|h|^2$ and $|\phi|^2$ on the super-Hubble scales as in Ref. [37].
In order to characterize the amplitude of tensor perturbations, one defines the tensor-to scalar-ratio,
\begin{equation}
   r(k=k_*=0.05~{\rm Mpc}^{-1}) = \frac{\Delta^2_{t*}}{\Delta^2_{s*}}
\,.
\label{r definition}
\end{equation}
Being equipped with these observable quantities, we find their newest constraint from Ref.~\cite{Ade:2015xua},
\begin{equation}
n_s =  0.9655 \pm 0.0062 \; (68 \%\; {\rm CL,\; Planck\; TT+lowP}, \alpha=0)
\,,
\label{observational limits: ns}
\end{equation}
Next we define the running of the spectral index,
\begin{equation}
   \alpha\equiv \bigg[\frac{dn_s(k)}{d\ln(k)}\bigg]_{k=k_*}
\,.
\label{running alpha}
\end{equation}
Ref.~\cite{Palanque-Delabrouille:2015pga} was able to show error-bars of $n_s=0.963\pm 0.0045$ and $\alpha=-0.0104\pm 0.0031$. These results should be confirmed in a future observation. Meanwhile, there is no direct measurement for tensor perturbations. Instead, the literature quotes upper bounds. For example, BICEP2/Keck and Planck Collaborations found~\cite{Ade:2015tva}
$r<0.12~(95\%{\rm CL})$, more recently~\cite{Array:2015xqh} BICEP2/Keck collaboration finds,
\begin{equation}
r<0.09\; (95\% {\rm CL},\; {\rm at}\; k_*=0.05~{\rm Mpc}^{-1})
\qquad ({\rm BICEP2/Keck})
\,.
\label{BICEP/Keck constraint on r}
\end{equation}
Future observations, in particular space missions, such as CoRE and LiteBird, will significantly improve the upper bound on tensor perturbations. In the following section, we will consider observable predictions of $R^p$ inflationary model as $p>2$, in which the corresponding value for $r$ is around $0.002$ as $n_s=0.965$.

\subsection{Cosmological perturbations in $R^p$ inflationary model}
Here, these observable quantities $n_s$, $r$ and its running index $\alpha$ are related to $R^p$ inflationary model in Einstein frame as it is well known that they are frame-independent \cite{Prokopec:2013zya}. In the light of Eqs.~(\ref{Einstein frame potential},~\ref{Fphi1}) and definition of $f(\Phi_E)$, the potential in Einstein frame can be explicitly written as,
\begin{equation}
V(\phi_E)=V_0e^{-2\sqrt{\frac{2}{3}}\frac{\phi_E}{M_p}}\left(e^{\sqrt{\frac{2}{3}}\frac{\phi_E}{M_p}}-1\right)^{\frac{p}{p-1}}
\,,
\label{the potential in Einstein frame}
\end{equation}
where $V_0=\frac{M_P^2}{2}(p-1)p^{p/(1-p)}\lambda^{1/(1-p)}$ and it agrees with \cite{Motohashi:2014tra}. This potential recovers $R^2$ inflation as $p=2$, for which $V(\phi)= \frac{3}{4}M^2M_P^2(1-e^{\sqrt{\frac{2}{3}}\frac{\phi}{M_p}}
)^2$ that was first proposed by Ref.~\cite{Barrow:1988xi}, $M$ is the energy scale which can be determined by the amplitude of the
observed power spectrum for primordial perturbations and $M\thickapprox10^{13}~{\rm GeV}$. In order to illustrate potential as $p>2$, we also reproduce the plot showed in Ref.~\cite{Motohashi:2014tra}.

\begin{figure}[h!]
 \centering
  \includegraphics[width=0.6\textwidth]{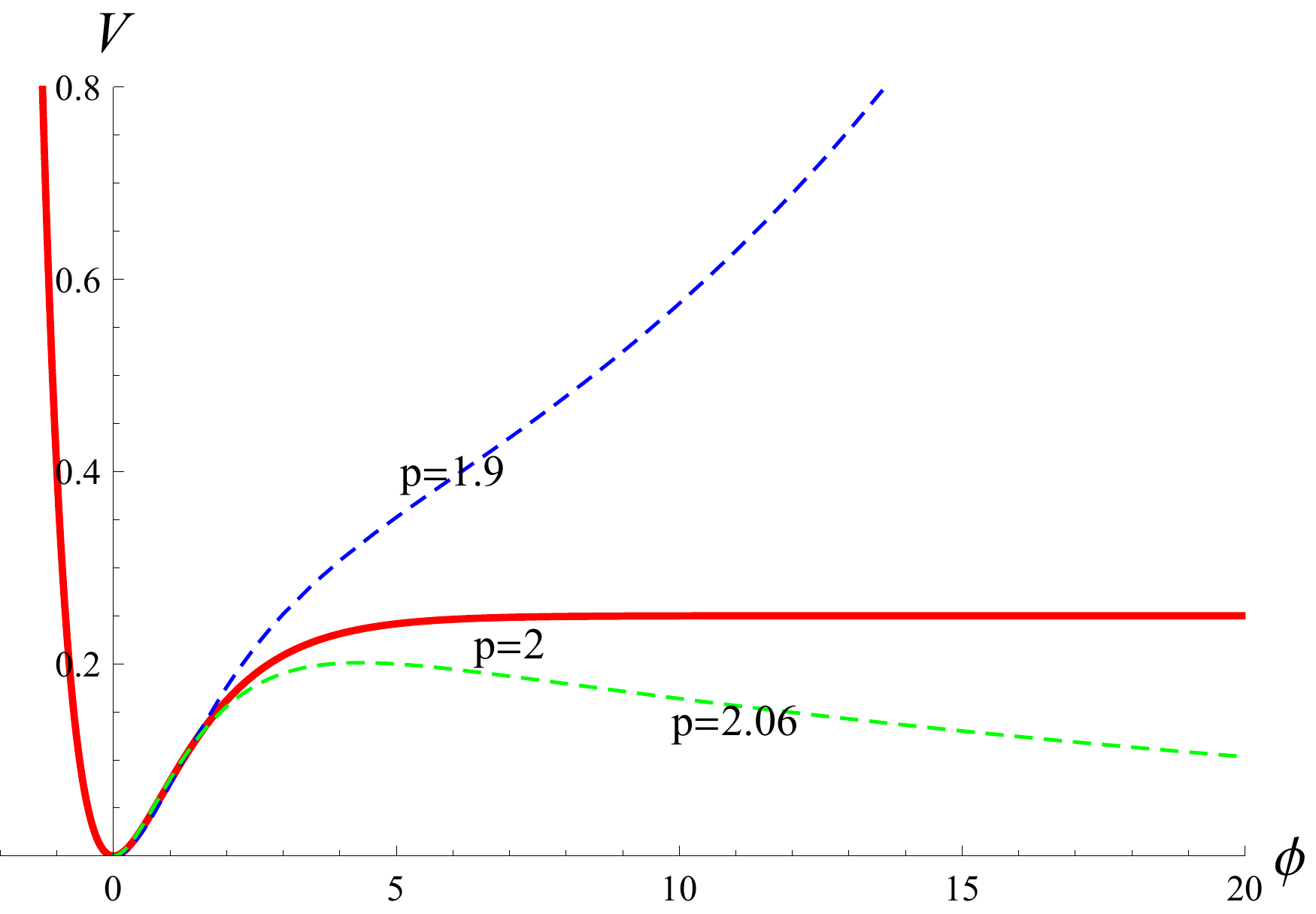}
 \caption{Potential for $R^P$ inflationary model with p=2(red thick line), p=1.9(blue dashed line)
   and p=2.06(green dashed line).}
\label{potential1}
 \end{figure}

Figure (\ref{potential1}) shows three cases of potential for $R^P$ inflation when $M_P=1$. For $p=2$, this potential recovers the well-known $R^2$ inflation potential. When $p\neq2$, the potential shows a deviation from $R^2$ inflation. As $p\geqslant 2$, there is a maximum value for $\phi_m^E=M_P\sqrt{\frac{3}{2}}\ln[\frac{2(p-1)}{p-2}]$. This scenario is quite different from $R^2$ inflation, there is a decay process for potential on the right of $\phi_m^E$. However, this process is not physical since Ricci increases as $\phi_E$ increases, as can be seen from Eq.~(12). As a consequence, the scalaron must shrink. Thus, we only need to consider the part that is left from $\phi_E$ for physical inflationary period.

In what follows we consider the scalar power spectral index $n_s$,
 tensor to scalar ratio $r$ and its running index $\alpha$.
Most inflationary models exhibit attractor behavior, which means that the physical parameters
 (tensor and scalar spectra) can be expressed in terms of the inflation amplitude individually,
 such as $\dot{\phi}_E$ is a function of $\phi_E$, this can be considered as so-called slow roll approximation.
 $\dot{\phi}_E$ $\ddot{\phi}_E$ are small in slow roll approximation and we expand in these parameters.
 Thus, we want to apply this slow roll attractor to our mdoel.
 From the canonical quantization of scalar and tensor perturbations (\ref{EOM mode function})
 and (\ref{observer spectra}), we can see that $n_s$ and $n_t$ can be expressed in this attractor regime in terms
 of geometric slow roll parameters,
\begin{eqnarray}
  n_s = 1-2\epsilon_E - \eta_E
\,,\qquad
  n_t = -2\epsilon_E
\,,\\ \epsilon_E = - \frac{\dot H_E}{H_E^2}
\,,\qquad \eta_E = \frac{\dot \epsilon_E}{H_E\epsilon_E}
\,.
\label{spectral index: geometric1}
\end{eqnarray}
The spectral index $n_s$ is a function of $k$, meanwhile its running $\alpha=\frac{dn_s}{d\ln k}$ can also expressed in terms of geometrical slow roll parameters,
\begin{equation}
   \alpha =-\eta_E(2\epsilon_E+\xi_E)\,,\qquad \xi_E\equiv \frac{\dot\eta_E}{\eta_E H_E}
\,.
\label{alpha in terms of slow roll paramater}
\end{equation}
Apparently, $\alpha$ also depends on $k$. Since current observations only constrain the upper limits of $\alpha$,
we can denote $\alpha$ by $\alpha(k_*)$. Based on slow-roll approximation, one can also express the potential slow roll
parameters in terms of the inflationary potential,
\begin{eqnarray}
 \epsilon_V &=& \frac{M_{\rm P}^2}{2}\left(\frac{V_E^\prime}{V_E}\right)^2
,\nonumber\\
  \eta_V &=& M_{\rm P}^2 \frac{V_E^{\prime\prime}}{V_E}
\,,
\nonumber\\
\xi_V^2 &=& M_{\rm P}^4\frac{V_E^\prime V_E^{\prime\prime\prime}}{V_E^2}
\,,
\label{traditional slow roll parameter}
\end{eqnarray}
where $V'_E=\frac{dV_E}{d\phi_E}$. Together with $d\ln(k)=d\ln(Ha)$ and Friedmann equation (\ref{Friedmann 2}), Eq.~(\ref{traditional slow roll parameter}) implies that,
\begin{equation}
\epsilon_V=\epsilon_E
\,,\qquad
\eta_E = -4\epsilon_V-2\eta_V
\,,
\label{spectral index: relation}
\end{equation}
and therefore
\begin{equation}
  n_s =1 -6\epsilon_V + 2\eta_V
\,,\qquad
  n_t= -2\epsilon_V
\,.
\label{spectral index: traditional 1}
\end{equation}
Furthermore, one can show that,
\begin{eqnarray}
  r &=& 16\epsilon_E = 16\epsilon_V = -8 n_t
\,\label{r in traditional slow roll}\\
  \alpha &=& 16\epsilon_V\eta_V-24\epsilon_V^2-2\xi_V^2
\,.
\label{alpha in terms of slow roll parameter}
\end{eqnarray}
Eq.~(\ref{r in traditional slow roll}) is the one field consistency relation which can test the validity of single field inflation. Notice that $n_s$, $r$ and $n_t$ are of first order in terms of slow roll parameters, while $\alpha$ is of second order in slow roll parameters and thus it is tiny. Our later calculation will confirm our discussion.
Finally, it is useful to define the number of e-foldings $N$, which in slow-roll approximation reads,
\begin{equation}
N\approx N_E=\int^{t_e}_{t}H_Edt_E\simeq \frac{1}{M_P^2}\int^{\phi_E^e}_{\phi_E}\frac{V_E}{V_E,_{\phi}}d\phi_E
\,.
\label{e folding number}
\end{equation}
where $t_e$ denotes the time at the end of inflation ($\epsilon_E=1$) and $V_E,_{\phi_E}=\frac{d V_E}{d \phi_E}$. Since Jordan frame is the physical frame, it is worth in investigating how large is the difference for the number of e-folding between Einstein frame and Jordan frame. One can easily obtain the following relation
between

\begin{equation}
H_J=\frac{\dot{\Omega}}{\Omega}+H_E
\,.
\label{relation for Hubble paramter}
\end{equation}
where $\dot{\Omega}=\frac{d\Omega}{dt}$, $H_J$ is Hubble parameter in Jordan frame. Using this result into Eq.~(\ref{e folding number}) and in the light of (\ref{Fphi1}), we can derive a relation between the e-folding number,
\begin{equation}
dN_E=dN_J+\frac{dF}{2F}
\,.
\label{relation for Hubble paramter}
\end{equation}
  Upon taking account of Eqs. (7) and (9) with (37), we get that the difference for e-folding numbers
in two frames is $-\sqrt{\frac23}\frac{d\phi_E}{M_P}$. Since $\dot\phi_E$ is small
in slow roll, we conclude that to good approximation,  $N_J\approx N_E$.

In the light of  Eqs.~(\ref{traditional slow roll parameter}),~(\ref{Einstein frame potential}),~(\ref{Fphi1}) and Eq.~(\ref{e folding number}), we can reproduce the most important observable quantities for $n_s$, $r$ and $\alpha$ from Ref.~\cite{Motohashi:2014tra},
 \begin{eqnarray}
\begin{aligned}
&n_s-1=-\frac{8(2-p)[(2-p)E_k^2+p(E_k-1)]}{3[2(p-1)E_k-p]^2},\\
&r=\frac{64E_k^2(2-p)^2}{3[2(p-1)E_k-p]^2},\\
&\alpha=-\frac{32p(2-p)^2E_k(E_k-1)(2E_k-3p+4)}{9[2(p-1)E_k-p]^4}
\,,
\label{observable quantities}
\end{aligned}
\end{eqnarray}
where $E_k=e^{4(2-p)N(\phi_E)/(3p)}$. Arming with these three observable quantities in terms of the e-folding number $N$ and $p$,
in what follows we discuss observational constraints on $n_s$, $r$ and $\alpha$.

\section{Results}

As the main result of this paper, we show how $n_s$, $r$ and $\alpha$ vary with the e-folding number $N_E$ and $p$. The range of e-folding number here is $50\leqslant N(\phi_E)\leqslant 60$.

\begin{figure}[h!]
 \centering
  \includegraphics[width=0.6\textwidth]{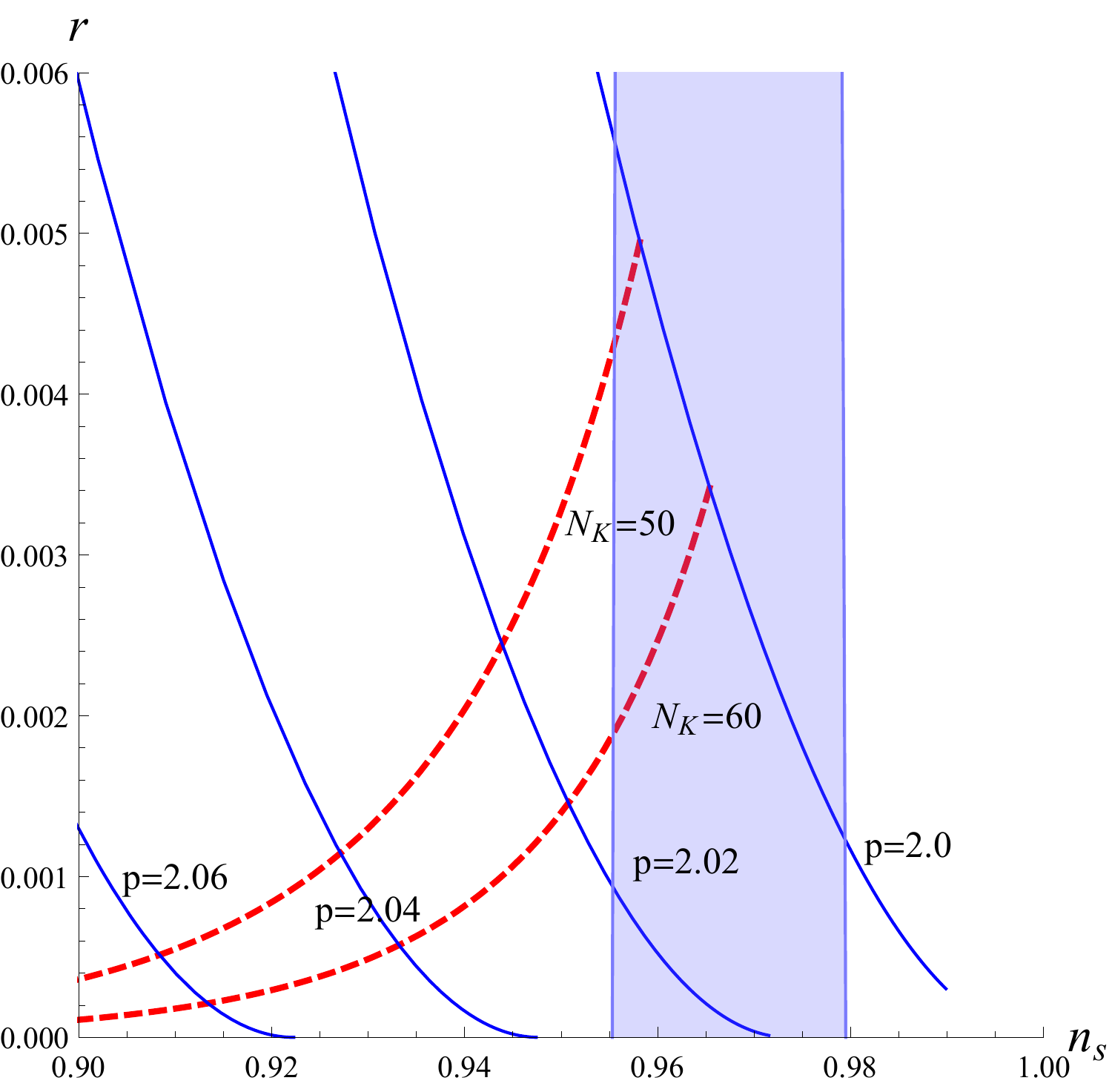}
 \caption{The tensor-to-scalar ratio $r$ as a function of the scalar spectra index $n_s$ for $p=2.06, 2.04, 2.02,2.0$ in blue solid lines. We show three curves for $N=50, 60$ in red dashed lines. The current upper limit on the tensor-to-scalar ratio (\ref{BICEP/Keck constraint on r}), $r<0.09$ is outside the plot's range. The range of $n_s$ from Eq.~(\ref{observational limits: ns}) is presented in blue shadow region.}
\label{nsr}
 \end{figure}

In figure 2, we show how the tensor-to-scalar ratio $r$ depends on $n_s$ as a function of $N$ and $p$. When $p$ approaches $2$, it recovers the $R^2$ inflation in which $r\simeq 0.003$ as $N\simeq 60$. There is a deviation from $R^2$ inflation varying with $p$ which shows $r$ decreases as $P$ enhances. By requring $N \in [50, 60]$, the case of $p\geq2.02$ in $R^p$ inflation can be ruled out. As for the allowed range for $r$, its magnitude lies from $0.001$ to $0.004$ within the reasonable range of $N$.

\begin{figure}[h!]
 \centering
  \includegraphics[width=0.6\textwidth]{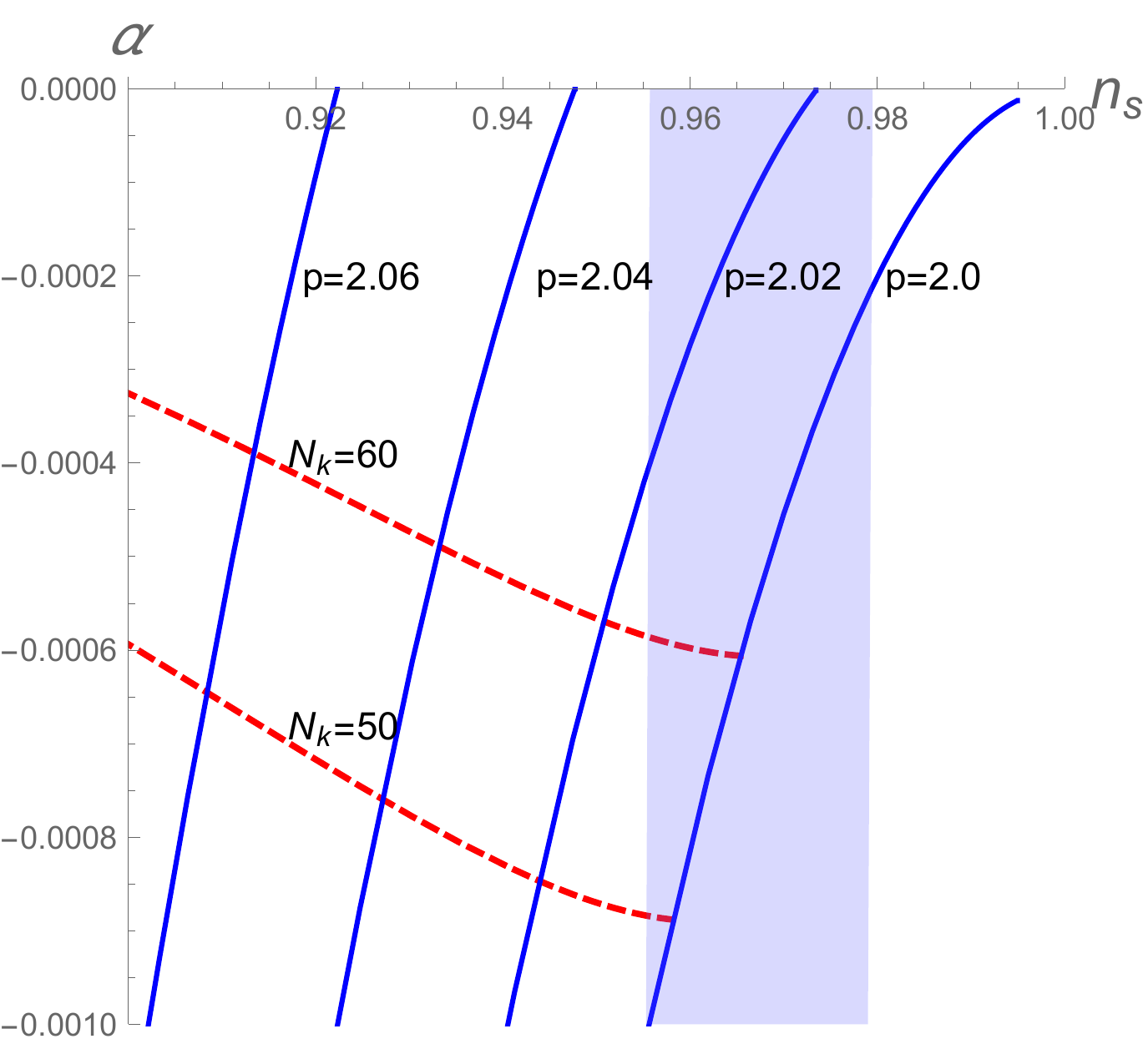}
 \caption{The running of scalar spectral index $\alpha$ as a function of the scalar spectral index $n_s$ for $p=2.06, 2.04, 2.02,2.0$ in blue solid lines. We show three curves for $N=50, 60$ in red dashed lines. The shadow shows the allowed values of $n_s$ according to (\ref{observational limits: ns}). The current Plank Collaboration limits on the running or scalar spectra index $\alpha \in [-0.01,0.004]$.}
\label{ns_alpha1}
 \end{figure}

In figure 3, we impose the identical constraints as showing in figure 2. We show how the scalar spectra index $n_s$ as a function of its running $\alpha$ varies with $p$ and $N$. As $p$ increases, the absolute value of $\alpha$ will decrease and this trend changes dramatically as $p$ grows above $2.02$. Thus, we can see that the validity of $R^p$ inflation is quite sensitive to $p$. On the other hand, when e-folding number is of valid range from $50$ to $60$ adopting the same constraints as in (\ref{observational limits: ns}), the corresponding value of $\alpha$ is within $[-0.0004,-0.0006]$. Thus, in order to detect $\alpha$ in our model, the measurement ought to be improved about one order of magnitude with respect to current observations, which will occur when the next generation of CMB space observatories will be launched in space (such as COrE). Nevertheless, together with a detection of $r$, an observation of $\alpha$ would play an important milestone in testing various models in a near future. For completeness, we also give the plot of $n_s-\alpha$. Due to lacking more accurate data of $\alpha$ and $r$, we only show that the possible observables range as expected. Another feature of figure 4 is that $r$ enhances as $p$ decreases.

\begin{figure}[h!]
 \centering
  \includegraphics[width=0.6\textwidth]{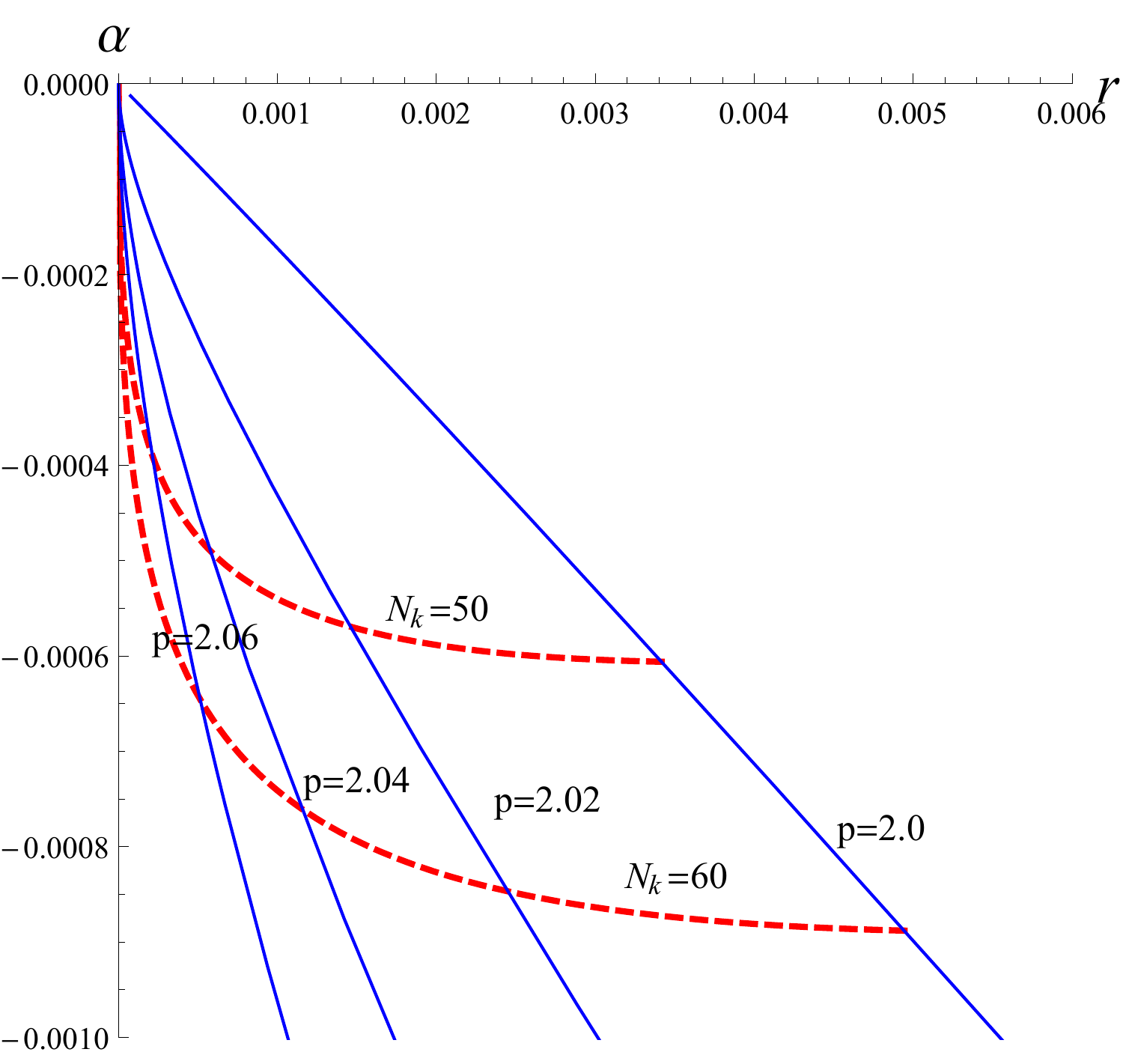}
 \caption{The running of scalar spectra index $\alpha$ as a function of the tensor-to-scalar ratio $r$ for $p=2.06, 2.04, 2.02,2.0$ in blue dashed lines. We show three curves for $N=50, 60$ in red solid lines. The shadow part show the allowed values of $n_s$ in (\ref{observational limits: ns}). The current upper limit on the tensor-to-scalar ratio (\ref{BICEP/Keck constraint on r}) $r<0.09$}
\label{r_alpha1}
 \end{figure}

To summarize, we have shown that the observable such as $n_s$, $r$ and $\alpha$ are strong sensitive with $p$ in $R^p$ inflationary model when $p\geq 2$. The model is ruled out as $p>2$ because of rather small values of $n_s$ as shown in Figure (\ref{nsr}). Due to its enhanced absolute value of $\alpha$, this class of inflationary model can be tested by future dedicated space CMB missions.

\section{Conclusion}

In this work, we analyse the $R^p$ inflationary model \cite{Motohashi:2014tra}, where $p$ is slightly larger than two. The effective potential for scalaron exhibits a local maximum. We show that there is inflation when scalaron rolls toward to the smaller value from the maximum as shown in Figure \ref{potential1}. One can also get inflation as the scalaron rolls towards the larger value, but it is not viable since it leads to a shrank universe. We perform an anlysis of observables in the model and show that this model is valid as $2\leq p\leq 2.02$ which is presenting in Figure \ref{nsr},~\ref{ns_alpha1} and \ref{r_alpha1}. When $p\geq2$, in particular, the scalar spectral index $n_s$ becomes smaller than what observations suggest which can be shown in Figure \ref{nsr}. Generally, as $p$ increases, $n_s$, $r$ and $|\alpha|$ decrease, the next generation of CMB observable may be challenging.

\section{ACKNOWLEDGMENTS\label{except}}
I am grateful for inspiring discussions with Prof. T. Prokopec. This work is in part supported by the D-ITP consortium, a program of the Netherlands Organization for Scientific
Research (NWO) that is funded by the Dutch Ministry of Education, Culture and Science (OCW). L.H Liu is funded by the
Chinese Scholarschip Council (CSC).


%

%




\end{document}